\documentclass[conference]{IEEEtran}

\usepackage[utf8]{inputenc} % allow utf-8 input
\usepackage[T1]{fontenc}    % use 8-bit T1 fonts
\usepackage[bookmarks=false]{hyperref}       % hyperlinks
\usepackage{url}            % simple URL typesetting
\usepackage{booktabs}       % professional-quality tables
\usepackage{amsfonts}       % blackboard math symbols
\usepackage{color} %pour mettre de la couleur
\usepackage[super]{nth}
\usepackage{calc}%     needed for the width/height calculations
\usepackage{float}
\usepackage{multirow}
\usepackage{amsmath}
\usepackage{algorithm2e}
\usepackage{algorithmic}
\usepackage{mathtools} % for DeclarePairedDelimiter
\usepackage{bm} %bold math
\usepackage{graphicx,subcaption}
\usepackage{tabularx}
\usepackage{bm}

\usepackage{tikz}
\definecolor{mycolor1}{rgb}{1,0.2,0.3}
\definecolor{mycolor2}{rgb}{0.2,0.3,1}
\definecolor{Gray}{gray}{0.9}
\definecolor{LightCyan}{rgb}{0.88,1,1}
\definecolor{Orange}{rgb}{1,0.5,0.2}
\definecolor{Yellow}{rgb}{1,1,0}
\definecolor{Green}{rgb}{0.5,0.9,0}
\definecolor{Blue}{rgb}{0,0.7,1}
\definecolor{LtBlue}{rgb}{0.4,0.8,1}
\definecolor{DkBlue}{rgb}{0., 0.447,0.69}
\definecolor{Purple}{rgb}{.7,0.5,1}

\tikzstyle{tau1} = [mycolor1, dashed]
\tikzstyle{tau2} = [mycolor2, densely dashdotted]
\tikzset{cross/.style={path picture={ 
  \draw[black]
(path picture bounding box.south east) -- (path picture bounding box.north west) (path picture bounding box.south west) -- (path picture bounding box.north east);
}}}

\newcommand{\eqdef}{\overset{\text{def}}{=}}

\DeclarePairedDelimiter\abs{\lvert}{\rvert}%
\DeclarePairedDelimiter\norm{\lVert}{\rVert}%
\makeatletter
\let\oldabs\abs
\def\abs{\@ifstar{\oldabs}{\oldabs*}}
\let\oldnorm\norm
\def\norm{\@ifstar{\oldnorm}{\oldnorm*}}
\makeatother

\usepackage{todonotes}

\newcommand{\isabelle}[1]{\todo[inline,color=orange!40]{#1 -- Isabelle}} 
\newcommand{\benjamin}[1]{\todo[inline,color=red!40]{#1 -- Benjamin}}

\newcommand{\toremove}[1]{\todo[inline,color=red!40]{to be removed \-- \\ #1 \\ \-- to be removed}}

\renewcommand{\toremove}[1]{}
\title{Anticipating contingengies in power grids\\ using fast neural net screening}

\author{\textbf{Benjamin Donnot$^\ddagger$ $^\dagger$}\thanks{Benjamin Donnot corresponding authors: benjamin.donnot@inria.fr}, \textbf{Isabelle Guyon$^\ddagger\bullet$, Marc Schoenauer$^\ddagger$}, \\ \textbf{Antoine Marot$^\dagger$, Patrick Panciatici$^\dagger$} \\
  $\ddagger$ UPSud and Inria TAU, Université Paris-Saclay, France. \\
 $\bullet$ ChaLearn, Berkeley, California.  $\dagger$ RTE France.}

\makeatletter
\newcommand\footnoteref[1]{\protected@xdef\@thefnmark{\ref{#1}}\@footnotemark}
\makeatother

\usepackage[xindy,toc]{glossaries}

\newglossaryentry{line}
{
  name=transmission line,
  description={is a generic term to denote everything that transmit flow in a power grid. Not to be confused with \gls{conn}}
}
\newglossaryentry{lf}
{
  name=load-flow,
  description={is both a model and an algorithm that allow people to compute the flow on each transmission line in a power grid given the injections. It can alternatively be called power-flow}
}
\newglossaryentry{inj}
{
  name=injection,
  description={is an object that inject (or extract) power from a \gls{grid}. Most often, if the value of power injected is positive we will speak of production. If this value is negative, it will be called a load, or consumption}
}
\newglossaryentry{grid}
{
  name=power grid,
  description={aims at transmitting the power from the production to the consumers. Be careful not to confuse it with \gls{ann} or graph}
}
\newglossaryentry{ann}
{
  name=neural network,
  long=artificial neural network,
  description={is a machine learning model. Be careful not to confuse it with \gls{grid} or graph}
}
\newglossaryentry{graph}
{
  name=graph,
  description={is a representation of a \gls{ann}. Not be confused with \gls{ann} or \gls{grid}}
}
\newglossaryentry{conn}
{
  name=connection,
  description={is weight in an \gls{ann}. Not to be confused with \gls{line}}
}
\newglossaryentry{topo}
{
  name=topology,
  description={corresponds to the way the objects (loads, generators, lines etc.) are connected together in an power-grid. Not to be confused with \gls{arch}}
}
\newglossaryentry{arch}
{
  name=architecture,
  description={applies only to neural networks, and corresponds to the way the connections are organized within. Not to be confused with \gls{topo}}
}
\newglossaryentry{reftopo}
{
  name=reference topology,
  description={topology of the power grid before any lines disconnection TODO}
}
\newglossaryentry{skeleton}
{
  name=underlying skeleton,
  description={complete set of connections TODO} 
}
\newglossaryentry{barch}
{
  name=base architecture,
  description={architecture with $\sigma_0$ for reference topology. TODO}
}
\newglossaryentry{mask}
{
  name=mask,
  description={connection pattern TODO}
}
\newglossaryentry{globtopo}
{
  name=global topology,
  description={list/vector of local topologies}
}
\newglossaryentry{loctopo}
{
  name=local topology,
  description={topology of a substation TODO}
}

\definecolor{mycolor1}{rgb}{1,0.2,0.3}
\definecolor{mycolor2}{rgb}{0.2,0.3,1}
\definecolor{mycolor3}{rgb}{0.,0.4,0.}

\newcommand{\distas}[1]{\mathbin{\overset{#1}{\kern\z@\sim}}}%
\newsavebox{\mybox}\newsavebox{\mysim}
\newcommand{\distras}[1]{%
  \savebox{\mybox}{\hbox{\kern3pt$\scriptstyle#1$\kern3pt}}%
  \savebox{\mysim}{\hbox{$\sim$}}%
  \mathbin{\overset{#1}{\kern\z@\resizebox{\wd\mybox}{\ht\mysim}{$\sim$}}}%
}

\begin{document}

\maketitle

%\vspace{-0.5cm}
\begin{abstract}
We address the problem of maintaining high voltage power transmission networks in security at all time. This requires that power flowing through all lines remain below a certain nominal thermal limit above which lines might melt, break or cause other damages. Current practices include enforcing the deterministic ``N-1'' reliability criterion, namely anticipating exceeding of thermal limit 
%(so-called ``contigency'')
for any eventual single line disconnection (whatever its cause may be) by running a slow, but accurate, physical grid simulator. New conceptual frameworks are calling for a probabilistic risk based security criterion and are in need of new methods to assess the risk. To tackle this difficult assessment,  we address in this paper the problem of rapidly ranking higher order contingencies including all pairs of line disconnections, to better prioritize simulations. We present a novel method based on neural networks, which ranks ``N-1'' and ``N-2'' contingencies in decreasing order of presumed severity. We demonstrate on a classical benchmark problem that the residual risk of contingencies decreases dramatically compared to considering solely all ``N-1'' cases, at no additional computational cost. We evaluate that our method scales up to power grids of the size of the French high voltage power grid (over $1~000$ power lines).
\end{abstract}

\section{Introduction}
% \benjamin{Maintenant que latex compile, re checker les notations, et verifier que la suite de papier a un sens avec ce qu'on raconte ici}
% \benjamin{Faut-il mieux avoir R(V) = sum (Z - V) ou R(V) = sum(V). Le second cas est plus "intuitif" on "découvre" du risque au fur et à mesure qu'on fait des simulations. Dans le premier cas par contre, on peut bien écrire "min R" (dans le second c'est "max R")}
% \antoine{ Introduire ceci dans le contexte d'une application industrielle proche du temps réel, contexte dans lequel il y a un compromis difficile à trouver entre exploitation et exploration. Notre budget computationnel est nécessairement fini mais notre risque est probablement maximum proche de l'échénace d'où une nécessité de continuer à explorer les risques mais de manière ciblée. Pour un budget donné, On cherche à maximiser le risque estimer par le vrai simulateur ou à minimiser le risque résiduel. }

%\benjamin{Faire un truc au début avec le vocabulaire: event = line outage = line contingencies = unplanned outage ||  bad event = line outage leading to at least one other line exceeding thermal limit}
%\benjamin{Apres on ne parle que de "event", plus de contingencies}
%\benjamin{Enlever le "ranking" des figures}
%\benjamin{sur le label de x, mettre c(nu)}
%\benjamin{sur la figure, c'est r max (x), pas juste r max}
%\benjamin{changer les bleus pour qu'il soit cohérent}

Although our application domain is power systems, the problem that we address is relatively general. It consists in ranking rapidly items/samples/entities, hereby referred to as ''events'' with a fast ``proxy'' method, then re-evaluating results with a slower (more accurate) method. The goal is to identify ``bad'' events, which should be ranked first. The hope is to gain in speed without sacrificing accuracy of the overall ranking or to increase the accuracy of ranking at no additional computational cost. In our power grid setting, an "event" will be "line disconnection(s)" or "line outage(s)" and will be referred to as a "contingency". 
Today's, ranking entails running a power grid simulator to compute power flow through all lines at a given time (so-called ``load flow''). ``Bad'' events or "dangerous contingencies" correspond to the disconnection of a power line, that put the grid in an unsecure state: the thermal limit of (at least) one other line is exceeded due to an excess of current flowing through it.
If no remedial actions are taken quickly, this overloaded line becomes out of service as well. As the grid weakens, the problem propagates like a snow ball effect (which may lead to a black-out). This can typically happen when the power grid is overloaded at peak consumption times and/or one or several lines have been disconnected. 
In this paper, we will consider that the system is in quasi-stationary conditions under given static ``injections'', corresponding to balanced productions and consumptions. This is typical of the conditions experienced by grid operators (dispatchers), which review grid conditions at intervals of five minutes. In our setting, grid events consist in line disconnections. Our novel contribution is to use neural networks to perform a first fast ranking of would-be ``bad contingencies'', using an architecture we recently published~\cite{donnot:hal-01695793}.
%\benjamin{Bad event = une ligne déconnectée qui créé des dépassements de limites thermiques. C'est pas ce qui est marqué dans le texte. Une contingencies = perdre une ligne, une "bad contingencies" / "dangerous contingencies" c'est perdre une ligne ET que de ce fait, une AUTRE ligne dépasse sa limite thermique.}

The problem of ranking contingencies in power system is not new. In \cite{vaahedi1999voltage} for example, the authors studied voltage stability problems after some unplanned line disconnection (contingency) occurs. 
%In \cite{abdulrazzaq2015contingency}, the authors address both the voltages and flows, but they don't use machine learning. For ranking the contingencies that can cause overflow, they use the "DC" approximation, a widely used approximation in the power system community. One limitation of their method is that they must evaluate all the contingencies using the slow high end simulator. 
%\isabelle{I removed a sentence because I do not see the difference. Can we compare ourselves to the DC approximation?}
%\benjamin{Non. Personne ne fait vraiment du screening en DC, et il me faudrait longtemps pour faire les simulations. DC c'est bien, mais en vrai personne l'utilise vraiment. On se compare aux opérateurs, c'est déjà pas mal non ?}
%The authors do not rank the contingencies, they rather ranks the contingencies that need to be addressed by corrective method from the most dangerous to the least one. The same kind of work can be also found in \cite{hussain2011} for example. 

Other authors use machine learning to address power system related problems. Most papers address the problem of classifying grid states according to given security criteria (\cite{wehenkel1997machine}, \cite{wehenkel2012automatic}, \cite{saeh2008static}, \cite{fliscounakis2013contingency}), or predicting how a system will react after an unplanned event occur (\cite{Duchesne2017}). To the best of our knowledge, using neural networks for screening ``bad contingencies'' has not been done before.
%Some other work \cite{saeh2008static} uses artificial neural networks, and compare it to a load flow simulator for a similar task. But again, the authors does not rank contingencies, only try to see if they will lead to overload or not. 

Closest to our work, \cite{sunitha2013online} use Artificial Neural Networks and Restricted Bolztman Machine to predict "Composite security indices", a concept similar to our scores $\hat{s}$ (see section \ref{sec:method}), allowing us to rank the contingencies $z$. A key difference between their work and ours is that we use our neural network to make predictions under multiple contingencies. Our neural network architecture is more efficient and better adapted to the task at hand.

Our method can also be seen as a particular case of "point-wise approach" by the "learning to rank community" (\cite{li2011short}). These models are widely used in documents retrieval (\cite{shashua2003ranking}) or medical drug discovery (\cite{doi:10.1021/ci9003865}). To our knowledge, none of the above methods applies directly to contingencies ranking in power system.

%the use the same test case as we do. On comparison, our ranking methodology rely on adapted for the task neural neural network, and are tested on data far from the training set, to have an idea on how the system will react facing a new, unseen, situation.
%\benjamin{TODO: citer le papier de Laurine sur le "coût" de la contingence pour enlever un des problèmes qu'on a }
%\cite{sidhu2000contingency}: use of artificial neural network

%\isabelle{We are missing the state of the art. Also search for "cite" to complete citations}
%\benjamin{On ne fait pas du "learning to rank" ici, mais quelques part, on montre qu'on n'a pas besoin (erreur presque parfaite dans le cas le plus favorable.}
\section{Statement of the problem and notations}

Formally, given a set $\mathcal{Z}$ of candidate events $z$, our goal is to identify the subset $\mathcal{B}$ of ``bad'' events by evaluating with a slow simulator the smallest possible number of events $z$. We run first a fast ``proxy'' simulator (our neural network), on all events in $\mathcal{Z}$ to obtain a set $\mathcal{V}$ of ``at risk'' candidates to be verified with the slow simulator. This saves us the effort of checking the remaining events in set $\mathcal{Z}-\mathcal{V}$. 

%Isabelle: the notation \mathcal{H} does not appear until much later in the paper, it is useless to introduce it here.

Under our assumptions, a ``system state'' $x$ consists of the power flowing in all lines, resulting from given (fixed) injections, for a specific grid topology. We always analyze a situation corresponding to a fixed state in this paper and sometimes omit $x$ for brevity of notation. We also omit to specify time ordering, although states are time ordered. What we denote by $z\in \mathcal{Z}$ are  {\bf sudden would-be (potentially disruptive) events}, corresponding to a change in grid topology, such as a {\bf line disconnection}, assuming injections remain constant. In some sense, $z$ is a variation $\Delta x$. In the language of power systems, such events are also referred to as {\bf ``contingencies''}. Therefore in this paper ``event'' and ``contingency'' will be synonymous and ``bad event'' and ``bad contingency'' will be synonymous.  An contingency $z$ might arise with probability $p(z)$\footnote{For instance, events $z$ might be single line disconnections occurring with probability $p(z)=\pi(1)$ or double line disconnections occurring with probability $p(z)=\pi(2)=\pi(1)^2$.} and is associated with a loss function $L(z;x)$. The overall risk is defined as\footnote{In the power system community this criterion is similar to the ``discarding'' principle of the GARPUR methodology (see \cite{karangelos2016probabilistic} Eq. 3, for instance).}:
\begin{equation}
R_{\max}(x) = \sum_{z \in \mathcal{Z}} p(z) L(z;x)
\end{equation}

%\benjamin{citer garpur ici et mieux introduire les différence de notations}

%\antoine{Mieux mettre en évidence que notre Loss L vaut 0 ou 1 ici. On a besoin de le retrouver pour l'équation 3 si on a lu rapidement et l'information est noyée ici}

\noindent where, in our application context, we assume that $L(z;x)$ is the $\{0, 1\}$ loss, with $0$ meaning that the contingency $z$ arising in state $x$ is innocuous and $1$ that it is risky or ``bad'' or "dangerous" for our system ({\em i.e.} at least one line, still in service after $z$ arose, will exceed its thermal limit)\footnote{This is a simplification. The real damage of the grid would endure after contingency $z$ would require computing a full "cascading failure" (as presented in \cite{1508.01775} for example), which is computationally too expensive to calculate presently, even for a small test case like ours.}. Thus:
\begin{equation}
  L(z;x) =  \left\{
    \begin{aligned}
      0 ~&~\text{``No current flowing on any line} \\
         &~\text{exceeds the line thermal limit } \\
         &~\text{under contingency $z$} \\
         &~\text{in grid state $x$''} \Rightarrow \text{OK}\\
      1 ~&~ \text{``Otherwise''} \Rightarrow \text{``Bad'' event}
    \end{aligned}
  \right.
\end{equation}
% \noindent  \\

Because of computational cost, we will only evaluate the subset $\mathcal{V} \subset \mathcal{X}$ of events, which will be identified as ``potentially bad'' by the fast ``proxy'' ({\em i.e.} neural network), using the slow (but accurate) simulator ({\em i.e.} load flow simulator). We assume that corrective actions are taken for all events found ``truly bad'' in $\mathcal{V}$ using the slow simulator, bringing the loss to zero for such events.

Thus, the \emph{residual} risk, corresponding to events in $\mathcal{Z} - \mathcal{V}$, which we neglected to evaluate with the slow simulator, is:
\begin{equation}
R(\mathcal{V}; x) = \sum_{z \in \mathcal{Z} \-- \mathcal{V}}   p(z) L(z;x)
\end{equation}

The residual risk $R$ is bounded between:
\begin{equation}
	R(\mathcal{Z}; x)=0 \text{~~~and~~~}   R(\emptyset; x)=R_{\max}(x)
\end{equation}

%\antoine{ En lecture rapide ce n'est pas très limpide de garder le même symbole R pour parler de risque et de risque résiduel. je mettrait par exemple un r cile ou un delta R pour parler du risque résiduel. à discuter
%}

The fast proxy provides a total ranking of $z \in \mathcal{Z}$ yielding nested subsets $\mathcal{V}_1 \subset \mathcal{V}_2 \subset \dots \subset \mathcal{Z}$ including an increasing number of events $z$, starting from those considered most at risk by the screening method. We can thus study $R(\mathcal{V}_i; x)$ as a function of $i=\text{rank}(z)=\abs{\mathcal{V}_i}$, where $\abs{\cdot}$ denotes the cardinal of a set.\\

In our simulations, the computational cost of the fast proxy is negligible compared to running the slow simulator. Previous experiments indicate that a $2000\times$ speed-up is achievable\footnote{Once the data $x$ are loaded, Hades2, the fast high end simulator used by RTE makes approximately $300$ ms to compute a load flow, first experiment on neural network show that a they can perform $5900$ load-flow per second, provided that all $6000$ grid states are already loaded in memory.} Hence the computational cost is proportional to the number of events $\abs{\mathcal{V}}$ identified ``potentially at risk'' by the neural network that are actually run by the simulator (no offset for the neural network computational cost):
%\benjamin{citer le technical paper, ou le papier ou on va mettre plein de détails sur les architectures, les datasets, et autres imap utilisées.}} Hence, the computational cost $C\mathcal{(V)}$ is proportional to the number of events $| \mathcal{V} |$ evaluated by the slow simulator. Denoting the maximum computational cost to evaluate all events under consideration $C\mathcal{(Z)}$ by $C_{max}$ and omitting index $i$, we have:

\begin{equation}
\frac{\text{rank}(z)}{| \mathcal{Z} |}
= \frac{|\mathcal{V}|}{| \mathcal{Z} |} 
= \frac{C(\mathcal{V})}{C_{max}}
\end{equation}

\begin{figure}[htp!]
\centering
  \includegraphics[width=0.95\linewidth]{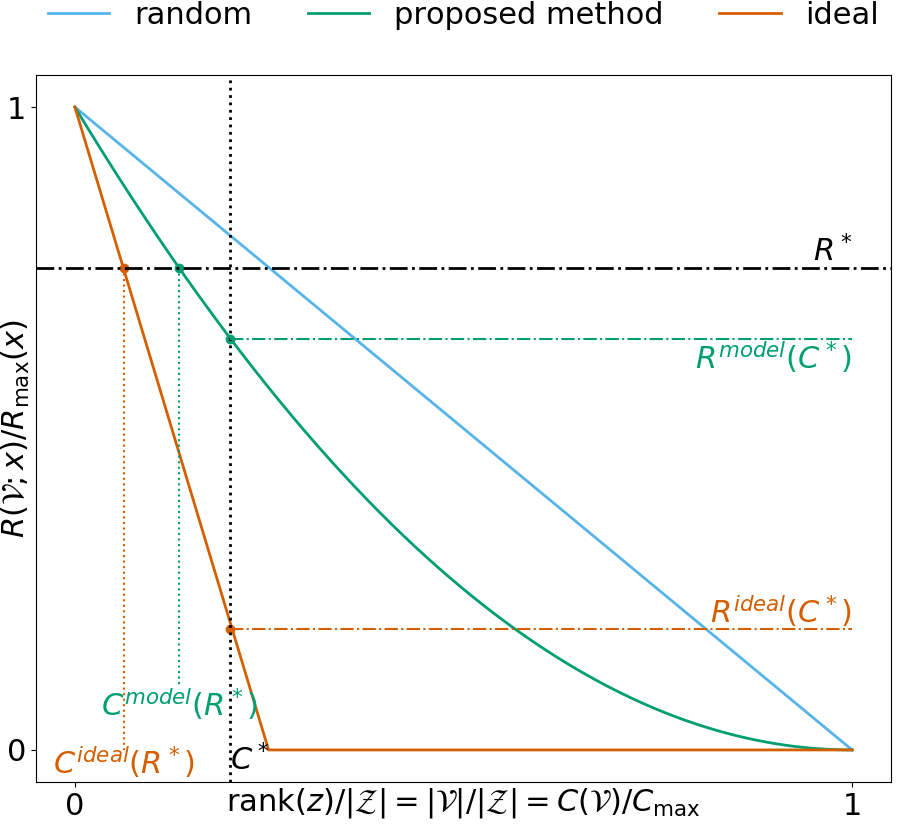}
  \caption{{\bf Didactic representation of risk as a function of computational cost}. We introduce notations in this schematic plot (not corresponding to actual numerical simulations).  The black horizontal dash-dotted line represent $R^*$: the maximum residual risk we are willing to accept. The Black vertical dotted line represents $C^*$ the maximum computational cost we are willing to pay. We assume for simplifity that all events have the same probablity, which yields staight segments for the blue and orange curves.}
  \label{fig:ideal_repr}
\end{figure}

In Figure \ref{fig:ideal_repr} we schematically represent a hypothetical case to illustrate our notations. It is the opposite of a lift curve used in marketing in which revenue = (1 - risk) is plotted as a function of investment. We consider a given state of the system $x$ and study the effect on the residual risk $R(\mathcal{V}; x)$ of various ranking strategies obtained by various screening methods ({\em i.e.} what we referred to as fast ``proxy''). The horizontal axis represents increasing size nested subsets $\mathcal{V}$ of events $z$ ``called bad'' by a given screening method. Such events incur a computational cost $C(\mathcal{V})$ to evaluate which of them are ``truly bad'' (using the slow simulator). To limit costs as much as possible, we want $R(\mathcal{V}; x)$ to decrease fast, {\em i.e.} we want $\mathcal{V}$ to contain as many ``truly bad'' events as possible. We represented three cases:
\begin{itemize}
\item The {\bf \color{Blue} blue line} is the expected value of the risk for a random ranking. 
\item The {\bf \color{Orange} orange curve} is the risk for an ``ideal'' ranking in which all ``bad'' events are ranked first. 
\item The {\bf \color{Green} green curve} is the ranking obtained by a given proposed screening method. 
\end{itemize}
To select between alternative methods, we can either fix a maximum budget $C^{*}$ and compare $R(\mathcal{V}; x)$ or set a maximum risk $R^{*}$ and compare $C(\mathcal{V})$.
%, or find an optimal trade-off between $R(\mathcal{V}; x)$ and $C(\mathcal{V})$, e.g. by optimizing $R(\mathcal{V}; x) + \lambda C(\mathcal{V})$ (for $\lambda>0$). This latest approach will not be envisaged in this paper.

It is worth noting that estimating $R(\mathcal{V}; x)$ is another difficult aspect of the problem. In this paper, because we use a benchmark of modest size, we exhaustively compute $L(z;x)$ for all $z$ with the slow simulator to draw curves such as those of Figure \ref{fig:ideal_repr}. In practice $R(\mathcal{V}; x)$ might have to be approximated by replacing $L(z;x)$ by an approximate loss $\hat{L}(z;x)$, obtained using power flows estimated by our ``proxy'' simulator. This yields a risk estimator:
\begin{equation}
\hat{R}(\mathcal{V}; x) = \sum_{z \in \mathcal{Z} \-- \mathcal{V}}   p(z) \hat{L}(z;x)
\end{equation}

%\isabelle{The lines C* and R* do not need to cross at point \{C*, R*\} in the general case}

\section{Aspects specific to the power grid problem}

Our proposed method relies on previously published work~ in \cite{bonnot:hal-01581719} and \cite{donnot:hal-01695793} in which we devised a neural-network trained with "guided dropout", which allows us to predict power flows in power grids for given topology variants while training only on a small subset of these. 

%\toadd{

%}

Our new contributions in this paper are (1) to formulate the detection of "bad" events ($=$ dangerous contingencies) as a ranking problem; (2) to use our ``guided dropout'' network as ``proxy'' simulator to provide a total ranking of events; and (3) to evaluate the performance of such ranking in several scenarios implied by slow or fast changing environments (distribution of the grid state $x$). In particular, we examine 2 variants of our ``proxy'' simulator: one in which training and test data are identically distributed (implying that either the distribution of $x$ does not change over time or that the proxy can be often retrained), and another one where contingencies never seen during training also need to be ranked (testing the ability of the method to deal with novelty in real time, a critical aspect for real time operation in a fast changing environment). %This last case is here to evaluate how the method behave in 

\iffalse
\benjamin{Ajout}
\textcolor{Purple}{
Our proposed method relies on previously published work~\cite{donnot:hal-01695793} and ~\cite{bonnot:hal-01581719} in which we devised neural-network trained with "guided dropout", which allows us to predict power flows in power grids with many topology variants while training only with a small subset of topology variants. Our new contribution in this paper is to use this technique to provide a total ranking of "at risk" situations both in the case where all topologies variant have been learned, or when only a subset of these have been seen during training. The first case corresponds to an operational process where a neural network have to be re-trained very often, so that it is always evaluate on data close to what it is trained on. On the contrary, the second case aims at mimicking the fact that a model is trained on demand, not very often.}
\benjamin{fin ajout}
\fi

Following the nomenclature of power systems, we recall that events correspond to line outages and are called ``contingencies''. "Bad" events correspond to contingencies after which at least one power line exceeds its ``thermal limit'' (a limit on maximum flow essentially set to prevent the line from melting). In this paper, we consider only two kinds of contingencies: ``single contingencies'', denoted by $z_i$, representing the disconnection of a single power line, and ``double contingencies'' $z_{i,j}$ representing the disconnection of two lines.

After the power grid suffers a single contingency, we will say its state is in "n-1" (notice the quotes, this is not an algebraic formula), namely: ``it has lost one in "n" lines''. If $n$ denotes the number of lines in our power grid, there are exactly $n$ different "n-1" grid states. Similarly, a power grid suffering a double contingency will be referred to as being in a degraded state "n-2". There are exactly $n(n-1)/2$ such "n-2" grid states.

To further elaborate on the specifics of our problem in application to power systems, we explain in the following subsections the security criteria used by TSO's (Transmission System Operator, responsible of the power grid safety) and define a baseline ranking method corresponding to the operational strategy in place in today's French power grid management.

%\subsection{Beyond "N-1" security policy}
\toremove{
%Power grid, like almost any system, tend to heat when they are crossed by electric current. In order to protect the equipment, and the people leaving nearby power line, there is, for each line in a power grid, a thermal limit, eg a maximum quantity of current allow to flow through it.

%Isabelle: most of these are repetitions
%Due to physical constrained, maximum flows are allowed to flow in a power line. This maximum flow can be referred as "thermal limit" mainly because it prevent power line from melting.

Commonly, TSO's operate the  grid using the so called "N-1" security policy. This policy stipulates that should \emph{ANY} unplanned single contingency occur, the flow on all the lines of the power grid must remain below their thermal limits. This terminology should not be confused with the "n-1" state in which the grid finds itself after one line disconnection. In fact assessing the "N-1" security requires computing $n$ load-flow, each one corresponding to one possible "n-1" grid state. This is a computationally intensive task. For example, the French power grid counts approximately $n \approx 10 000$ power lines. Thus, assessing the "N-1" security of this network requires  $\approx 10 000$ computations. In this context, it is understandable that TSO's do not operate under higher order security policies, such as "N-2" (two line disconnections), "N-3", and so on.
Our primarily motivation for studying "N-2" grid safety is that TSO operators must anticipate future grid states on an ever longer horizon to guarantee security. 
Indeed, the complexity of power grids in operation is constantly increasing, due to the integration of renewable energies, the interaction with foreign markets, and the presence of increasingly complex automata on the grid. In this context, remedial actions cannot always be taken in real time. Thus our events $z$ include leaps in the future, at a scale where a succession of single contingencies might happen fast enough that they cannot be repaired before another one hits the grid.
In this scenario, the probability for  $z$ to yield a degraded "n-2" state will be much higher in the future than it is today. Our work anticipates the need to consider multiple contingencies.  
}
\subsection{Risk modeling}
\begin{figure}[htp!]
\centering
  \includegraphics[width=0.95\linewidth]{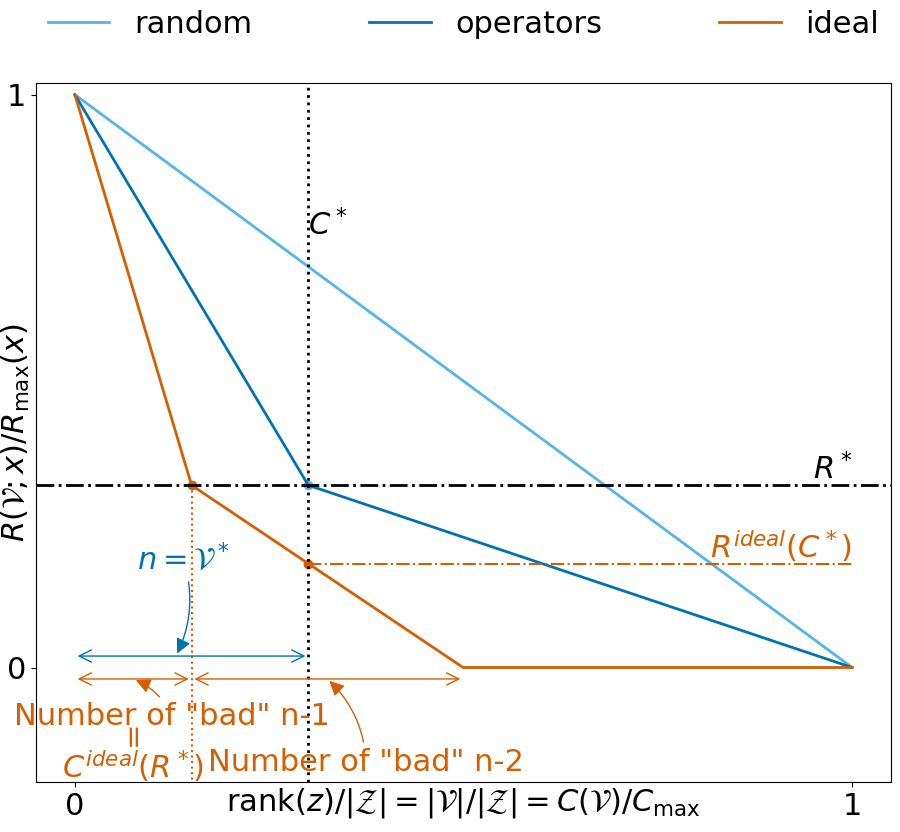}
  \caption{{\bf Application-specific risk as a function of cost} (didactic representation). We schematically illustrate the random curve (dark blue), baseline curve (light blue) simulating the "n-1" contingencies first, and the optimal curve (orange) simulating first all "bad" events.}
  \label{fig:ideal_representation_withop}
\end{figure}

In this paper, we focus on assessing the "N-2" security, that is the state of the system ("secure" / "non secure") after each possible "n-2" OR "n-1". Computing this safety would require $n(n+1)/2$ calls to the high end power grid simulator, since $n$ computations are required for all "n-1" contingencies (loss of one line) and $n(n-1)/2$ for all "n-2" contingencies (loss of two lines). In our notations, this means that the set of all events under consideration is:
\begin{equation}
\mathcal{Z}=\text{"N-2"}~~~\text{with}~~~\abs{\mathcal{Z}}=\frac{n(n+1)}{2}.
\end{equation}

We ignore the effect (and the residual risk) associated with higher order contingencies ("n-3", "n-4", etc.). As a further simplification, we assume that all single disconnections $z_i$ have equal probability:
  	\begin{equation} p(z_i) \eqdef \pi_{(1)}\end{equation}
and all double disconnection $z_{i,j}$ have equal probability:
  	\begin{equation} p(z_{i,j}) \eqdef \pi_{(2)}  = \pi_{(1)}^2. \label{eq:proba_emp}
    \end{equation}
In reality, such probabilities vary depending on factors such as line length, pair of line proximity, local climate, weather variations, etc. 

To illustrate the conceptual framework and our notations, we represent in Figure\ref{fig:ideal_representation_withop} a schematic hypothetical case. Our goal is to improve over the "N-1" policy, thus our {\bf baseline method} (represented as a {\bf \color{DkBlue} dark blue} line) will be to rank first all "n-1" contingencies. We define:
\begin{equation}
\mathcal{V^*}=\text{"N-1"}~~~\text{with}~~~|\mathcal{V^*}|=n,
\end{equation}
corresponding to an associated cost and residual risk:
\begin{equation}
C^*=C(\mathcal{V^*})~~~\text{and}~~~R^*=R(\mathcal{V^*},x).
\end{equation}
The dark blue curve is composed of two straight segments. It corresponds to the expected value of $R(\mathcal{V},x)$ when  ranking first (in random order) all "n-1" events, then (in random order) all "n-2" events. 
Between $|\mathcal{V}|=0$ and $|\mathcal{V}|=|\mathcal{V^*}|=n$, we have:
\begin{equation}
	R(\mathcal{V},x) = R_{\max}(x)-\alpha_1 \pi_{(1)} |\mathcal{V}|
\end{equation}
and between  $|\mathcal{V}|=|\mathcal{V^*}|=n$ and  $|\mathcal{V}|=|\mathcal{Z}|$, we have:
\begin{equation}
R(\mathcal{V},x) =  R(\mathcal{V^*},x) - \alpha_2 \pi_{(2)} |\mathcal{V}|
\end{equation}
where $\alpha_1$ and $\alpha_2$ are the fraction of ``bad'' "n-1" events (contingencies) and ``bad'' "n-2" events respectively.

To compare our proposed method with the baseline, we will either:
\begin{itemize}
\item Evaluate the {\bf reduction of residual risk} $R^*-R(\mathcal{V},x)$ for $C(\mathcal{V})=C^*$, $C^*$ being the {\bf maximum (computational) cost} we are willing to incur, or;
\item Evaluate the  {\bf reduction of cost} $C^*-C(\mathcal{V})$ for $R(\mathcal{V},x)=R^*$, $R^*$ being the {\bf maximum residual risk} we are willing to incur.
\end{itemize}
Our proposed method can at best achieve an optimal strategy ({\bf \color{Orange} orange  curve} in Figure\ref{fig:ideal_representation_withop}) in which all ``bad'' events (first all "n-1" then all "n-2" contingencies) are ranked first. This yields a first segment with slope $ -\pi_{(1)}$ as a function of $|\mathcal{V}|$, then a second segment with slope $ -\pi_{(2)}$.

\subsection{Parameters setting}
We wanted our simulations to be as close as possible to the situation of the real French power system.

Expert dispatchers (TSO operators responsible to operate the grid in security) estimate that a full "N-1" simulation yields approximately $100$ "bad" events (dangerous contingencies). This means that approximately $1\%$ of the single events should present a serious risk requiring a corrective action. To respect the order of magnitude of this proportion of "bad" events, we used a calibration dataset, which allowed us to \emph{set} the thermal limits\footnote{\label{supplemental}See supplemental material at \href{https://github.com/BDonnot/acpgunn}{https://github.com/BDonnot/acpgunn} for a detailed value of the thermal limits used.} of each line in our test case grid. Having set these values, we evaluate them on the full "N-1" for $100$ different grid states\footnote{See the section \ref{sec:data} for a detailed description of this dataset.}, requiring to compute $18~600$ load flows. Among all the "n-1" events investigated, $1.75\%$ were found unsafe in our simulations and $4.22\%$ for "n-2" events. \\

To be as close as possible to the French power system, we also calibrated the failure probabilities $\pi_{(1)}$ and $\pi_{(2)}$. We wanted that the ratio between the risk due to single failures and the one due to double failure to be constant across grid sizes, this leads to consider:
\begin{align}
    \pi_{(1)} & \approx 5.4 ~ 10^{-3} \\
    \pi_{(2)} & \approx 2.9 ~ 10^{-5}
\end{align}
\noindent For more information, the section V of the supplemental material\footnoteref{supplemental} shows in detail how we obtained these probabilities.

\section{Proposed methodology}
\label{sec:method}
In this section, we explain how we rank the contingencies $z$ with respect to their estimated severity $\hat{L}(z;x)$ for a given system state $x$.  $\hat{L}(z;x)$ is provided by an artificial neural network, trained on simulated data generated using a high-end load flow simulator.

Consider a fixed grid state $x$ and a given contingency $z$, we denote by $f_i$ the flow, computed with the high-end simulator, on the $i^{\text{nth}}$ line of grid $x$ after contingency $z$ occurs, and by $\bar{f}_i$ the thermal limit for this line.

We propose to first train a neural network with ``guided dropout'', as describe in \cite{donnot:hal-01695793} to approximate rapidly the power flow for the given grid state $x$ and contingency $z$. We will denote by $\hat{f}_i$ the flow predicted by our proxy (in this case our neural network) for the $i^{\text{nth}}$ line of the power grid.

It has been observed that neural network tend to be "over confident" in their predictions (see for example \cite{DBLP:journals/corr/NguyenYC14}). This overconfidence could lead to a bad ranking in practice with dramatic effects. We propose to calibrate the score of our neural network taken into account a fixed (yet calibrated) uncertainty by assuming:
\begin{equation}
	\forall i, (f_i-\hat{f}_i)  \sim \mathcal{N}\left(0, \sigma_i\right)
	\label{eq:error_normality}
\end{equation}
\noindent where $\sigma_i$  represents the model uncertainty for line $i$. We calibrate the vector $\bm{\sigma}$ (of dimension $n$) using a calibration set distinct from the training set the neural network was trained with and also distinct from the final test set we use to evaluate performance. On this calibration set, we compute the true values $f_i$, using the high-end simulator, and the predictions $\hat{f}_i$ coming from our proxy (neural network). Then, $\sigma_i$ is set to:
\begin{equation}
  \sigma_i \eqdef \frac{1}{\text{number of simulations}} . \sum_{\text{simulation} s} (\hat{f}_i - f_i)^2
\end{equation}

These $\sigma_i$'s are then used to compute the scores $\hat{L}_i$ that a given line is above its thermal limit as:
\begin{equation}
  \hat{L}_i \eqdef 1-F_{\sigma_i}(\bar{f}_i-\hat{f}_i)
\end{equation}
\noindent where $F_{\sigma_i}$ is the cumulative density function of the Normal law with mean $0$ and variance $\sigma_i$.

This gives us a score for each power line. For our problem, a grid is said to be ``non secure'' after contingency $z$, if at least one of its line is above its thermal limit. The score of the power grid, in state $x$ after contingency $z$, is then obtain with:
\begin{equation}
	\hat{L}(z; x) ~\eqdef~ \max_{1\leq i \leq n}  \hat{L}_i(z;x)
\end{equation}

We now have defined $\hat{L}(z;x)$. If all contingencies $z$ were equiprobable ($\pi_{(1)} = \pi_{(2)}$), we could use this approximation for ranking. In practice, some events occur more often that others and therefore must be prioritized accordingly. 

The first thing we tried was to rank directly the contingencies $z$ with respect to their relative cost $p(z) \hat{L}(z; x)$. This worked, but this order tends to be really conservative: all the single contingencies were ranked first. This can be explained. There is a $10^3$ relative factor between $\pi_{(1)}$ and $\pi_{(2)}$, so for a double contingency $z_{i,j}$ to be simulated before a single one $z_i$, this would mean that $\hat{L}(z_{i,j};x) > 10^3.\hat{L}(z_{i};x)$. But, we made various hypothesis, in particular that the error on the flow where normally distributed (see equation \ref{eq:error_normality}) which is not the case in reality. In general we have then $\hat{L}(z_i;x) - L(z_i;x) >> 10^{-3}$. Even if the contingency $z_i$ is harmless ($L(z_i;x) = 0$), this would imply $\hat{L}(z_i;x) > 10^{-3}$, so this single contingency would be ranked before any "n-2" contingency, which can be different in reality.

To remedy this problem, we tried various weighting schemes, and found that scaling the ``n-1'' event relatively to the ``n-2'' events with the logarithm of $\pi_{(1)}/\pi_{(2)}$ leads to the best results in almost every situation $(z;x)$.

To wrap up, all the grid states and contingencies are sorted with respect to their "scores" $\hat{s}(z;x)$ defined as:
\begin{equation}
\hat{s}(z;x) \eqdef 
    \left\{
	\begin{aligned}
          \hat{L}(z;x) \times \log\left(\frac{\pi_{(1)}}{\pi_{(2)}}\right) &\text{if } z \text{ is a single contingency} \\
          \hat{L}(z;x) &\text{ otherwise}
	\end{aligned}
    \right.
\end{equation}

\iffalse
    Note that the score $\hat{s}(z;x)$ should only be used for ranking, not to estimate the risk, since our estimate $\hat{L}(z; x)$ is biased:
    \begin{equation}
        \mathbb{E}\left( \hat{L}(z; x) \right) \neq p(z).L(z;x)
    \end{equation}
    \isabelle{what is $p(z)$?}
    \benjamin{cf equation 3. p(z) c'est toujours la meme chose.}

    \benjamin{C'est du au fait qu'on prenne un max sur 186 "nombres censé être distribués selon des lois normales". Et ce max a donc une valeur, non nulle. Je sais comment enlever ce biais (bootstraping sur le calibration dataset par exemple), mais vu que j'ai passé beauocup de temps à faire des figures propres, je n'ai pas implémenté ce truc. A voir si j'ai le temps lundi ou mardi...}
\fi

\section{Experiments}
In this section we report extensite experiments validating our proposed methodology, using a standard grid benchmark (the $118$ bus grid of the matpower library  \cite{Zimmerman11matpowersteadystate}, used for example in \cite{sunitha2013online}). We begin by explaining our data generation methodology to create a dataset with over 2 million samples. Then we report results showing a 50 fold speedup to achieve the same residual risk currently obtained with the ``N-1'' policy. We also show how the remaining computational time can be exploited to significantly reduce the residual risk by simulating higher order contingencies (for example double contingencies).

%In this section, we explain how we generated an extensive dataset for a commonly used benchmark grid with $186$ lines. We calculated the full ``N-1'' and ``N-2'' for calibration and test data ($75$ and $25$ grid states respectively) and the full ``N-1'' and a large number of ``n-2'' cases for $500$ different grid states for training (several million rows in our dataset). Using these calculations as ground truth we conducted extensive numerical experiments to evaluate our method, based on a neural network trained with ``guided dropout''~\cite{donnot:hal-01695793} 
%\benjamin{TODO citation ici}
%as explained in the previous section. We also evaluated how our method would scale to the full French Extra High Voltage power grid with over $3.000$ lines.

\subsection{Datasets generation}
\label{sec:data}
We conducted systematic experiments on small size benchmark grid (the $118$ bus grid) from Matpower \cite{Zimmerman11matpowersteadystate}, a library commonly used to test power system algorithms \cite{alsac1974optimal}. %We report results on the largest case studied: a 118-node grid with $n=186$ lines. 

We generated $500$ different grid states varying the injections $x$ of the initial grid given in Matpower. We used our knowledge of the French grid to mimic the spatio-temporal behavior of real data~\cite{donnot:hal-01695793} when simulating our new datasets. For example, we enforced spatial correlations of productions and consumptions and mimicked production fluctuations, which are sometimes disconnected. Target values were obtained by computing resulting flows in all lines with the proprietary \emph{AC} power flow simulator \emph{Hades2}.
 
{\bf Training and validation sets.} On these $500$ test cases, we then computed, still using the high-end simulator Hades2, the full "N-1" (making $500 \times 186 = 93.000$ load flow computations). We also simulated $k=5 000$ load flows randomly selected among the $\frac{186\times 185}{2} = 17~205$ possible "n-2" double contingencies. This last dataset counts then $2~500~000$ rows. We split this dataset in two subsets: $75\%$ for training and $25\%$ for validation/model selection {\em i.e.} for finding the best architecture and hyper-parameters of the neural networks. The detailed architecture tested can be found in the supplemental material at \href{https://github.com/BDonnot/acpgunn}{https://github.com/BDonnot/acpgunn}.
%\isabelle{don't forget to remove the TODOs}
%\benjamin{Bien vu! Je les avais mis pour me rappeler de mettre des truc dans le supplemental material, que je vais rédiger ce soir.}

{\bf Calibration and test sets.}  For the calibration dataset, used to fit the error of the model $\bm{\sigma}$, we simulated $75$ different grid states $x$, and the full "N-1" and "N-2" for all of these simulations. The test set is composed of $25$ grid states, and their full "N-1" (all single contingencies) and "N-2" (all double contingencies) associated dataset.

\subsection{Neural network architecture}
In this part, we will explain more about the neural network architecture used for the experiments, the way we trained it as well as the method used to find the so called "meta parameters" that gives the best results.

The used architecture (introduced in \cite{donnot:hal-01695793}) allows fast assessment of power flows given injections (productions and consumptions). We used this architecture that we called "guided dropout" as a fast proxy of the power flows simulator. The main idea of this purposely designed architecture is to adapt the \emph{architecture} of the neural network to the \emph{topology} of the grid. For example, we switch on/off hidden unit of the neural network depending on the sate off a given line in the power grid (connected / disconnected). We showed in this previous paper, that this architecture allows to super-generalize, eg. even when trained on a small grid configurations, it is able to predict flows properly to unseen grid states. 

To present more this architecture we consider the neural network is a function $F$ with input $x$, $l$ layers each counting $u$ units, and output $y$ (the flows ins our cases), and non linearities $\phi(\cdot)$. We have, for layer $k \in \{ 1, \dots l\}$:
\begin{equation}
 	y^{(k)} = \phi(W^{(k)}.x^{(k)})
    \label{eq:one_layer}
\end{equation}
\noindent where $y^{(k)}$ is the output of layer $k$ and $x^{(k)}$ its input ($x^{(k)} = y^{(k-1)}$), and $.$ denotes the matrix multiplication. For a sake of simplicity, bias are not written. The "guided dropout" (see \cite{donnot:hal-01695793}) consist in "masking" some part of the output of certain layers depending on external condition (here the state of a line: connected / disconnected). 

In our power system example, we decided to "guided mask out" the layer $k$ of our neural network. We then define (once and for all) for each power line $i$ a masking vector $\bm{m}^{(i)}$ of 0's and 1's. The equation for layer $k$ would become:
%two vectors $\bm{m}^{(1)}$ and $\bm{m}^{(2)}$ of 0's and 1's , with the condition if for a components $i$, $\bm{m}^{(1)}_i=0$ then for the same component $i$, $\bm{m}^{(2)}_i=1$. The equation for layer $k$ would become:
\begin{equation}
 	y^{(k)} = \bm{m} \odot \phi(W.x^{(k)})
\end{equation}
\noindent with $\odot$ denote the adamar product (element wise multiplication), and $\bm m$ is a mask build from the  $\bm{m}^{(i)}$'s with:
\begin{equation}
    \bm m = \left\{
        \begin{aligned}
        \text{identity} & \text{ if both $li_1$ and $li_2$ are connected} \\
        \bm{m}^{(1)}  & \text{ if $li_1$ is disconnected and $li_2$ is connected}  \\
        \bm{m}^{(2)}  & \text{ if $li_2$ is disconnected and $li_1$ is connected} \\
        \bm{m}^{(1)} \odot \bm{m}^{(2)}  & \text{ if both $li_1$ and $li_2$ are disconnected} 
        \end{aligned}
    \right.
\end{equation}

This can be viewed as an adaptation of the neural network configuration, depending on the power grid topology. The "presence / absence" of power line will have a direct impact on the "presence / absence" of unit in some layers.

To train the neural networks, we use the "adam" optimizer, variant of the stochastic gradient descent first introduced in \cite{kingma2014adam}. All the neural networks have been trained with the same number of minibatches. This number of minibatches have been calibrated such that the error on the validation set stop decreasing. Among all the meta parameters, we use a cross validation scheme to find the best combination of learning rate, number of layer and number of units per layer. The "relu" (rectifier linear unit) was used as the non linearity, except in the last layer. More details about the architecture used and the meta parameters can be found in the supplemental material at \href{https://github.com/BDonnot/acpgunn}{https://github.com/BDonnot/acpgunn}. 

%Recall that in this paper, we wanted to assess the feasibility of screening dangerous contingencies in power systems as fast a possible. To do that, we use one specific neural network architecture called "Guided Dropout" that were introduce in \ref{donnot:hal-01695793}, 
We trained the same architecture according to two schemes. The first one \textbf{"Guided Dropout (trained n-1 only)}" is trained only with single contingencies. During the training we never show input/output pairs coming from a power grid having suffered 2 contingencies. We do that to evaluate the robustness to the method to the input distribution: how the model will generalize facing unseen situations. In the second training scheme refered as "\textbf{"Guided Dropout}", we suppose that the distribution of the training set is really close to the one of the test set. This is a viable hypothesis in practice if the neural network can be retrained very often. \\

%We used all 187 variants of grid topologies with zero or one disconnected line ({\bf ``n-1'' dataset}) and randomly sampled 200 cases of pairs of disconnected lines ({\bf ``n-2'' dataset}), out of $186*185/2$. Training and test data were obtained by generating for each topology considered 10,000 input vectors (including active and reactive injections). To generate semi-realistic data,  we used our knowledge of the French grid, to mimic the spatio-temporal behavior of real data~\cite{bonnot:hal-01581719}. For example, we enforced spatial correlations of productions and consumptions and mimicked production fluctuations, which are sometimes disconnected for maintenance or economical reasons. Target values were then obtained by computing resulting flows in all lines with the AC power flow simulator Hades2,\footnote{A freeware version of Hades2 is available at \url{http://www.rte.itesla-pst.org/}}. This resulted in a ``n-1'' dataset of $1,870,000$ samples (we include in the ``n-1'' dataset samples for the reference topology) and a ``n-2'' dataset of $2,000,000$ samples. We used $50\%$ of the  ``n-1'' dataset for training, $25\%$ for hyper-parameter selection, and $25\%$ for testing. %All ``n-2'' data were used solely for testing. 
%\subsection{N-1 only}
%\benjamin{Mon super truc qui calcule tout et le met sur la figure fait de la merde avec le n-1, a voir plus tard}

\subsection{Results}
\begin{figure*}[htp!]
\centering
  \includegraphics[width=0.73\linewidth]{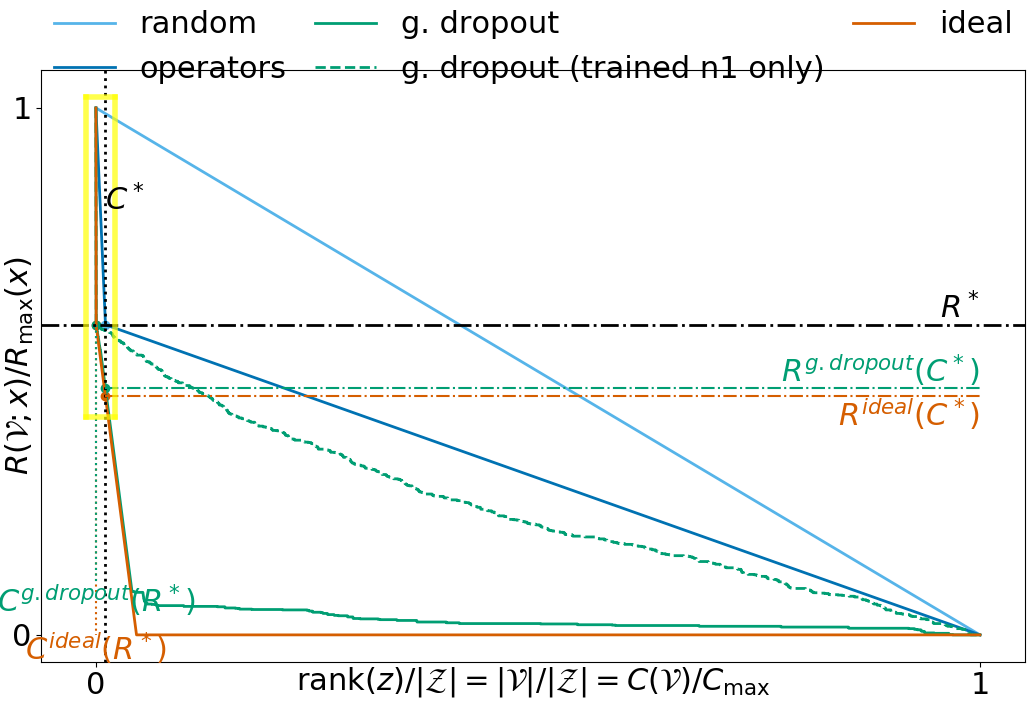}
  \caption{{\bf Contingency ranking using two neural networks}. The first network (plain line) is trained with both ``n-1'' and ``n-2'' contingencies, and it is tested on a situation not seen during training. The second neural network (dashed green line) is trained only on the "n-1" data: it has seen during training only "single contingencies" (a small fraction of all the contingencies it is tested on).}% 
  \label{fig:N2train_all}
\end{figure*}

\begin{figure*}[htp!]
	\centering
  \includegraphics[width=0.73\linewidth]{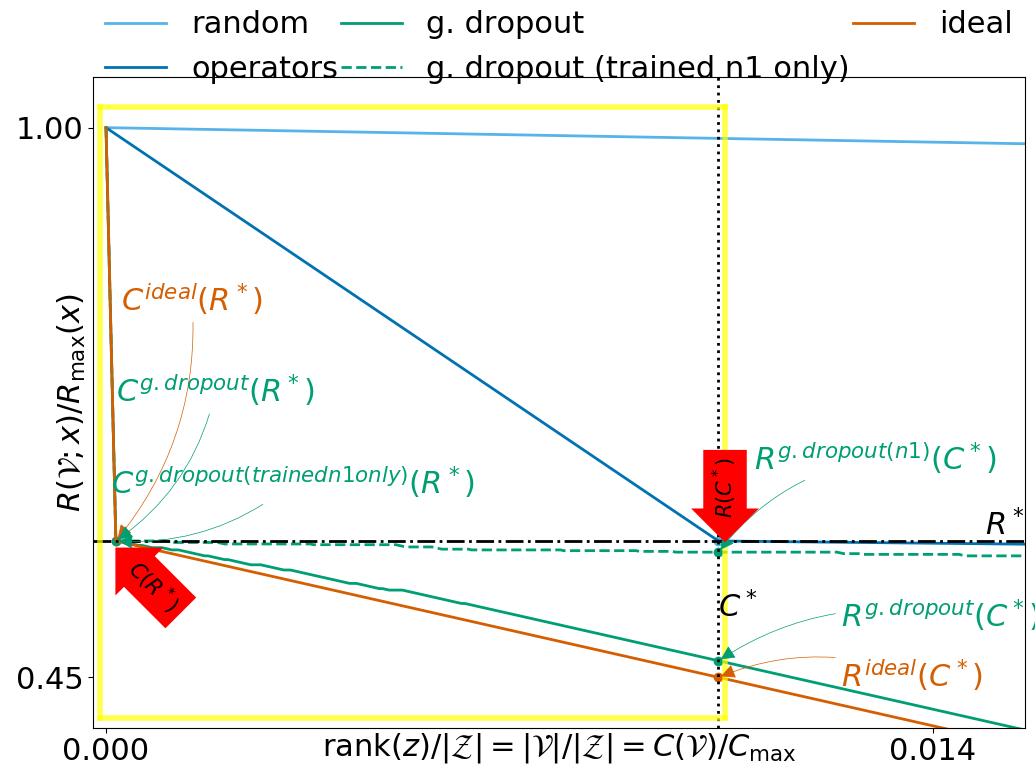}
  \caption{{\bf Zooming on the contingency ranking}. This figure corresponds to zooming the yellow rectangle in the previous figure, focusing on the region between $|\mathcal{V}|=0$ and $|\mathcal{V}|=n$ in which we hope to improve over the current ``N-1'' TSO security policy.}
  \label{fig:N2train_n1}
\end{figure*}

Figures \ref{fig:N2train_all} and \ref{fig:N2train_n1} represent the lift curves including:
\begin{itemize}
    \item the random ranking (light blue curve);
    \item the baseline ranking (dark blue curve) corresponding to ranking the ``n-1'' events first, as TSO operators would do;
    \item the ideal ranking (orange curve) in which the ``bad'' ``n-1'' events come first followed by the ``bad'' ``n-2'' events, and;
    \item the ranking for two different neural networks (green curves). 
  	\begin{itemize}
  		\item the plain curve represents the ranking for a neural network trained on the entire training set, which includes all "n-1" and randomly selected "n-2" contingencies; training and test data are similarly distributed.
        \item the dashed curve represents the ranking for a neural network trained on the "N-1" dataset only {\em i.e.} no double contingency is used for training. This neural network therefore exhibits ``super-generalization'' to multiple contingencies. Such change in distribution of data may occur in actual grid operation situations and the performances obtained are encouraging. 
	\end{itemize}
\end{itemize}
\noindent The curves represented in  Figure \ref{fig:N2train_all}  are computed for a grid state $x$ corresponding to a worse case scenario, yielding the highest loads of our dataset. This simulates ``peak demand'', when loads are $\simeq 40\%$ higher than for the reference case. Studying such extreme cases validates best our approach.
%We compare the results of two neural networks. The plain green line corresponds to training the neural network on the whole set $\mathcal{Z}$. The dashed line ????

%\isabelle{you are not computing all ``n-2'' so this should be a small n. Also as far as I understand, there is a single x and not the sample contingencies are seen during training and testing, or I am confused.}
%\benjamin{There is one single "x", and all N-1 and N-2 contingencies. So the situation (in term of injections) is new for sure, but not necessarily the contingencies. N-2 contingencies are completely new for the green dashed line model.}
%\benjamin{Concerning the small vs big "n", I think it is a big one here. This the "N-2" dataset, eg the dataset with all "n-2"...}

As we can observe in Figure \ref{fig:N2train_all}, both neural networks perform better than the emulation of TSO operator ranking in every situation. Remarkably, even the neural network that has never seen double contingencies during training is capable of ranking all single AND double contingencies quite well. This is an encouraging result indicating that our methods (called "guided dropout") is capable of ``super-generalization'' a form of transfer learning to cases of contingencies never seen before.

\begin{table}[ht!]

\caption{{\bf Comparison of methods} on the whole test set (mean over all $x$ cinsidered  $\pm$ one standard deviation). Gini coefficient (between 0 and 1, 1 is best), residual risk $R(\mathcal{V}^*)$ for a maximum computational budget $\mathcal{V}^*$and  computation cost $C(R^*)$ ensuring that the risk residual risk remains below $R^*$ (in test data). For both $R(\mathcal{V}^*)$ and  $C(R^*)$, smallest is best.}
\resizebox{\linewidth}{!}{%
\begin{tabular}{|lccc|c|}
\hline
            &   Operators &   \begin{tabular}{c} G. Dropout \\ {\tiny trained n-1 only} \end{tabular} &   G. Dropout &   Ideal \\
\hline
 Gini coeff.    &         
\begin{tabular}{c} $ 0.41 $ \\ $ {\tiny \pm 0.04 }$ \end{tabular} &                             \begin{tabular}{c} $ 0.52 $ \\ $ {\tiny \pm 0.04 }$ \end{tabular} & 
\begin{tabular}{c} $ 0.95 $ \\ $ {\tiny \pm 0.01 }$ \end{tabular} & 
1.00 \\
\hline
$\frac{R(\mathcal V*)}{R_{\max}}$ &       
\begin{tabular}{c} $ 0.59 $ \\ $ {\tiny \pm 0.04 }$ \end{tabular} &     
\begin{tabular}{c} $ 0.58 $ \\ $ {\tiny \pm 0.04 }$ \end{tabular} &   
\begin{tabular}{c} $ 0.46 $ \\ $ {\tiny \pm 0.03 }$ \end{tabular} & 
\begin{tabular}{c} $ 0.44 $ \\ $ {\tiny \pm 0.03 }$ \end{tabular}  \\
\hline
 $C(R^*)$         &      
186 &    
 \begin{tabular}{c} $ 3   $ \\ $ {\tiny \pm 2 }$ \end{tabular} &       
 \begin{tabular}{c} $ 3   $ \\ $ {\tiny \pm 2 }$ \end{tabular} &   
 \begin{tabular}{c} $ 3   $ \\ $ {\tiny \pm 2 }$ \end{tabular} \\
\hline
\end{tabular}
}
%\isabelle{Table captions are usually on top. Maybe normalize the results (with Cmax and Rmax). -> pour $C$ ça donne des nombres trèes petits, donc je trouve ça plus simple de parler en terme de "budget" = nombre d'appel au simulateur lent. Pour $R$ c'est normalisé}

\label{tab:resu_allsitu}
\end{table}
%\isabelle{average over what? if this is an average, why is there no stdev?}
%\benjamin{Average over all the x's of the test set (25 different x's), no stddev because I didn't see the point, but why not -> c'est une stddev sur "x", c'est a dire que formellement c'est l'état de réseau "x" qui est "stochastique", un peu bizarre dans ce papier...}

%Another fact that should be noted is the gain in term of computational cost. In fact, we can reduce by a factor of $50$ the necessary budget to have a residual risk bellow $R^*$, the "N-1" risk today taken by operators.
The table \ref{tab:resu_allsitu} presents the Gini coefficient (normalized area under lift curve; an index between 0 and 1, highest is best)\cite{gini_paper}
%isabelle{in the absence of definition, we need a citation}

as well as the residual risk at a fixed computational budget $R(\mathcal V*)$, and the cost for achieving a residual risk below $R^*$. 
%\isabelle{les gens savent lire, ils faut commenter les resultats, pas les lire}
%For the neural networks that have seen every type of contingencies ("G. Dropout"; \nth{4} column of table \ref{tab:resu_allsitu}), this index is on average of $0.95$. It is of $0.52$ for the model that have never seen double contingencies during training. Both these numbers should be compare to the gini index of the operator ranking of $0.42$. 
Our neural network method is always better that the TSO operator strategy, with respect to all metrics.
The most promising results of Table \ref{tab:resu_allsitu} is in the \nth{3} row. Our neural networks perform as well as the optimal strategy. On average, we need to simulate only $3-4$ contingencies to achieve a residual risk below the risk taken by operators. This is at the sale of this problem a speedup by a factor of more than $50$. Given that we scale the ratio of "bad" contingencies to be representative of the French power grid, we expected speed-up in the same order of magnitude ($10$ fold improvement) for the French Extra High Voltage powergrid. 
%\isabelle{we need to roughly evaluate here how we anticipate that this will scale to the French grid}
We can achieve that with both neural networks, even the network trained only on ``n-1'' data! Testing on new contingencies only improve results, as seen on  Figure \ref{fig:N2train_all}. Thus, the time saved with our method can be used to further reduce the residual risk by investigating more candidate ``contingencies''.

To sum up, both neural networks are capable of correctly ranking first all  "bad" single contingencies, and single contingencies are ranked before double contingencies. 
%: $R^{g. dropout}(C^*)$ and $R^{g. dropout}(C^*)$ are equal to $R^{ideal}(C^*)$. 
%Isabelle: not sure that this notation is helpful.
%The proposed method scores does not seem to suffer from overconfidence.
%Isabelle: not sure what this over-confidence refers to.
%the scoring of "bad" "N-2" contingencies is never too high, which could lead to ranking a "non bad" double contingencies before a "bad" single contingencies. 
As in every ranking problem, there is a tradeoff between the hit rate (detecting true ``bad'' contingencies) and the false alarm rate (falsely diagnosing harmless contigencies as ``bad''). From the point of view of ensuring the security of a power grid, the severity of both types of errors is not symmetric. It is far more important to have a high hit rate than a low false alarm rate. The Gini coefficient does not capture such imbalance but $C(R^*)$ does.
What is practically important to ensure the adoption of the method by the power system community is that the residual risk curve $R(\mathcal V)$ decreases very fast, ensuring that the ``hit rate'' be high initially. 
Remarkably, both neural networks rank first the ``bad'' ``n-1'' contingencies., then chose among the worst ``n-2'' contingencies. The neural networks privileged a high``hit rate'' over lowering the ``false alarm rate'', a risk adverse behavior, which is expected to operate power systems in security.

%The ranking coming from the model trained only on "N-1" simulates almost all the single contingencies $z_{i}$ first and then starts to simulate the double contingencies $z_{i,j}$, only when it is "sure" that no single contingencies $z_i$ will lead to a "bad" situation. This behavior is quite encouraging. The neural network does not become too confident, and even though it performs poorly, this method does not make bad mistakes, such as not simulating a risky single contingency $z_i$. It prefers to be "risk averse", even if it leads to a poor amelioration of the risk. This is an expected behavior in such a critical system.

\section{Conclusion}
In this paper, we proposed a novel approach to rank the most dangerous contingencies in power grids (namely those leading to lines exceeding their thermal limits and posing problems of grid security). This approach is quite general: it can rank single contingencies (one line disconnection) or multiple contingencies (two or more lines disconnected), with different probabilities of occurring. In this paper we studies the case of single and double contingencies.

Our methodology in a nutshell can be summarized as follows: (1) Train a neural network to mimic a load flow simulator. (2) Use it (on new test data) to evaluate how close each line it to its thermal limits.(3) Rank contingencies accordingly in decreasing order of severity. Our simulations  results on a standard benchmark case are quite promising:

First, with this approach, we can accurately identify the most dangerous contingencies. In all cases, we could use a computational budget of only $2\%$ of the ``N-1'' policy budget (the strategy presently delyed in production).

Second, this method is robust to changes in distribution between training and test time. Training a neural network with only $0.5\%$ of all possible contingencies (single contingencies) is enough to achieve a better performance than today's "N-1" criterion.

We evaluated that this method is scalable to the full Ultra High Voltage powergrid, and may result in speed-ups by a factor of $2~000$ or more in power flow calculations.
%\isabelle{``We show in previous paper that training a neural network on this grid is realistic, and a $2~000$ speed up can be achieve in practice for computing flows'' Do not put this in the conclusion: some solid argmument must be made in the body of the paper}

Although our approach was developed for power systems and demonstrated on a particular case study, it is rather generic and could easily be extended to other domains in which risk can be evaluated by a high-end but slow simulator and simulations must be prioritized for efficiency reasons. 

\bibliographystyle{IEEEtran}
\bibliography{references.bib} 

\end{document}